\documentclass[
 aip,
 amsmath,amssymb,
 reprint,
]{revtex4-1}

\usepackage{graphicx}
\usepackage{subcaption}
\usepackage{amsmath}
\usepackage{dcolumn}
\usepackage{bm}

\usepackage[utf8]{inputenc}
\usepackage[T1]{fontenc}
\usepackage{mathptmx}
\usepackage{etoolbox}
\usepackage{xcolor}
\usepackage{soul}

\makeatletter
\def\@email#1#2{
 \endgroup
 \patchcmd{\titleblock@produce}
  {\frontmatter@RRAPformat}
  {\frontmatter@RRAPformat{\produce@RRAP{*#1\href{mailto:#2}{#2}}}\frontmatter@RRAPformat}
  {}{}
}
\makeatother
\begin{document}

\title{Superconducting Diode Effect in Two-dimensional Topological Insulator Edges and Josephson Junctions}

\author{H. Huang}
 \affiliation{Department of Physics and Astronomy, Purdue University, West Lafayette, Indiana 47907 USA}

\author{T. de Picoli}
 \affiliation{Department of Physics and Astronomy, Purdue University, West Lafayette, Indiana 47907 USA}

 \author{J.~I.~V\"ayrynen}
 \affiliation{Department of Physics and Astronomy, Purdue University, West Lafayette, Indiana 47907 USA}

\date{\today} 

\begin{abstract}
The superconducting diode effect -- the dependence of critical current on its direction --  can arise from the simultaneous breaking of inversion and time-reversal symmetry in a superconductor and has gained interest for its potential applications in superconducting electronics.
In this letter, we study the effect in a two-dimensional topological insulator (2D TI) in both a uniform geometry as well as in a long Josephson junction. 
We show that in the presence of Zeeman fields, a circulating edge current enables a large non-reciprocity of the critical current. We find a maximum diode efficiency 1 for the uniform 2D TI and $(\sqrt{2} - 1)^2 \approx 0.17$ for the long Josephson junction. 
\end{abstract}

\maketitle

In a superconductor with broken inversion and time-reversal symmetry, the critical current can depend on the direction of the current bias, an effect known as the superconducting diode effect (SDE)~\cite{2023NatRP...5..558N,jiang2022superconducting}. 
Although the phenomenon is not new~\cite{PhysRevB.49.9244,harrington2009practical,vodolazov2018peculiar}, the subject has attracted renewed interest after 
the suggestion that the effect 
 might be intrinsic to some  materials~\cite{ando2020observation,wakatsuki2017nonreciprocal,yuan2021supercurrent,wu2021realization,lin2021zero,daido2022intrinsic,ilic2022theory,diez2021magnetic,baumgartner2022supercurrent,Mazur2022,baumgartner2022effect,PhysRevB.105.104508,PhysRevB.106.205206,PhysRevB.106.224509,legg2022superconducting,PhysRevB.107.224518,PhysRevB.109.064511,PhysRevB.103.L060503,PhysRevLett.121.026601,2023arXiv230101881B,2023arXiv231111087Y,bauriedl2022supercurrent} as opposed to of extrinsic geometric nature ~\cite{2023Sci...382.1422Z,PhysRevB.49.9244,PhysRevB.72.172508,2022arXiv220509276H,burlakov2014superconducting,sundaresh21,2023arXiv230101881B,Gutfreund_2023,Golod_2022,davydova2022universal,Chiles_2023}. 
The effect can be realized in critical current of both uniform (junctionless) systems~\cite{PhysRevB.49.9244,vodolazov2018peculiar,ando2020observation,burlakov2014superconducting,wakatsuki2017nonreciprocal,PhysRevLett.121.026601,2022arXiv220509276H,lin2021zero,scammell2021theory,sundaresh21,bauriedl2022supercurrent,Gutfreund_2023,PhysRevLett.132.046003,harrington2009practical,yuan2021supercurrent,daido2022intrinsic,ilic2022theory,PhysRevB.106.205206,legg2022superconducting,PhysRevB.107.224518,PhysRevB.109.064511} as well as Josephson junctions~\cite{wu2021realization,PhysRevB.98.075430,diez2021magnetic,baumgartner2022supercurrent,Mazur2022,baumgartner2022effect,Golod_2022,PhysRevLett.131.096001,davydova2022universal,kokkeler2022field,zhang2021general,PhysRevB.103.L060503,2022arXiv221001037H,2023arXiv230101881B,Chiles_2023,PhysRevB.105.104508}, but the relation of the two diode effects has not been completely clear. 

A widely studied mechanism for inversion symmetry breaking in an intrinsic SDE is strong spin-orbit coupling~\cite{yuan2021supercurrent,daido2022intrinsic,ilic2022theory,2022arXiv221001037H,legg2022superconducting,PhysRevB.87.100506,PhysRevB.93.174502,PhysRevB.103.L060503,PhysRevLett.131.096001,PhysRevB.107.224518,2023arXiv231111087Y}. 
A prime example displaying the effects of strong spin-orbit coupling is the time-reversal symmetric two-dimensional topological insulator (2D TI)~\cite{doi:10.1126/science.1133734,kane_quantum_2005,PhysRevLett.95.146802,qian2014quantum}. 
These systems have been also studied in the superconducting state, see for example HgTe~\cite{hart2014induced,hart2017controlled,mandal2024magnetically} or monolayer 1T' WTe$_{2}$~\cite{sajadi2018gateinducedSC,fatemi2018electrically,2020NatPh..16..526L}.

Diode effect and related non-reciprocal effects have been theoretically proposed in 2D TI Josephson junctions~\cite{PhysRevB.92.035428,PhysRevB.98.075430,PhysRevB.108.L161403,Vigliotti_2023,PhysRevB.109.075412}, mostly in the short junction limit (junction length less than superconducting coherence length). 
In this paper, we show that a \textit{uniform}  2D TI edge can display a large diode effect in its critical current, when the current is limited by Cooper pair depairing~\cite{PhysRevB.107.224518}. 
We then show an analogous Josephson diode effect in a long Josephson junction. 
We attribute both diode effects to the same mechanism that is unique to 2D TI edge: a circulating current generated by  Zeeman field or magnetization.

\begin{figure}
    \centering
    \includegraphics{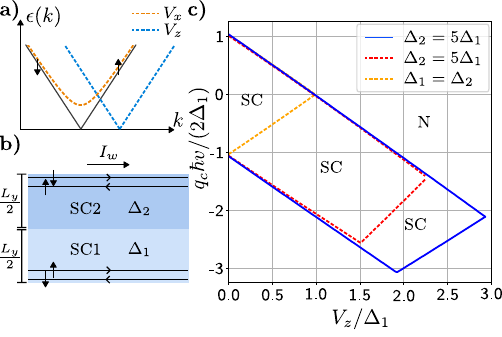}
    \caption{Superconducting diode effect in the critical current of a uniform (junctionless) 2D topological insulator ribbon. a)~Effect of a Zeeman field on the single edge dispersion. A field perpendicular to the spin quantization axis (which we choose to be $z$ axis) opens a gap in the dispersion, while a field along the spin quantization axis shifts the Dirac point momentum. b)~Geometry of the ribbon in our numerical model. The superconducting gap of the lower and upper half of the ribbon are $\Delta_{1}$ and $\Delta_2$, respectively, to break inversion symmetry. c)~Critical momenta $q_c^\pm $ versus $V_z$ for the ribbon, where only  the $V_z\geq 0$ part is shown since $q_c^\pm(V_z)=-q_c^\mp(-V_z)$. 
    The dashed lines are obtained for $\Delta_1=\Delta$ and the solid line is for $\Delta_1=\Delta/2$. For $\Delta_2=5\Delta_1$, we achieve diode efficiency $\eta=1$ at $V_z=\Delta_1$ as $q_c^+$ vanishes. 
    Along the solid boundary, the quasiparticle gap closes on one of the edges of the system.  
    In the case of larger gap,  $\Delta_1=\Delta,\Delta_2=5\Delta_1$ (the red dashed line), the size of the superconducting phase is limited by bulk states:    
    the positive-slope boundary is due to gapless bulk states as opposed to edge states in other cases. 
    Our numerical results indicate that $|q_c^\pm(V_z=0)| \approx 2 \Delta_1 / (\hbar v)$ is determined by 
 the Dirac velocity  $v$, thus approximately independent of $\mu$. 
    }
    \label{fig:Wire}
\end{figure}

Two-dimensional topological insulator has an insulating bulk gap but gapless helical (spin-momentum locked) one-dimensional boundary modes.  
The low-energy Hamiltonian for a single helical edge state is 
\begin{equation}
    H_\text{helical edge} = v \hat{p}_x \sigma_z + \sum_{i=x,y,z} V_i \sigma_i, \label{eq:HwireNormal}
\end{equation}
where $v$ is the edge mode Dirac velocity,  $\hat{p}_x = -i \hbar \partial_x$ is the momentum operator and $\sigma_{x,y,z}$ the spin Pauli matrices; we choose the helical edge spin quantization axis to be the $z$-axis. 
We also include a magnetic perturbation $V_i$ that breaks time-reversal symmetry of the edge and allows the system to present the diode effect.
The term $V_i$ may arise from Zeeman coupling to magnetic field as $V_i = \frac{1}{2} g \mu_B B_i$ or from exchange coupling to a magnetic material; for concreteness we call this a Zeeman field in the rest of the article. 
A perpendicular Zeeman field $V_i$ (e.g., in the $x$-direction)    opens a gap in the edge dispersion, 
whereas a  field along the spin quantization axis, $V_z$, merely polarizes the edge and shifts the Dirac point position, 
see Fig.~\ref{fig:Wire}a. 
In a finite system  $V_z$ will therefore induce a circulating current along the boundary of the system~\cite{PhysRevB.108.L161403}. This current will lead to various non-reciprocal effects, previously mostly studied in Josephson junctions~\cite{PhysRevB.83.220511,PhysRevB.92.035428}. 
Here, we first show a large superconducting diode effect in the absence of a junction and then discuss diode effect in a long Josephson junction. 

Because the edge Hamiltonian consists of counterpropagating states with opposite spins, conventional singlet s-wave superconductivity can be induced. 
In the presence of a supercurrent and a Zeeman field along $z$, the (2nd quantized) Hamiltonian   of a single proximitized 2D TI edge  is given by 
$\mathcal{H} = \frac{1}{2} \int \! dx \, \Psi^\dagger (x) H(x,q)  \Psi(x)\,,$
\begin{equation}
H(x,q)=(v  \hat{p}_x\sigma_{z}-\mu)\tau_{z}+V_{z}\sigma_{z}+\Delta e^{iqx\tau_{z}}\tau_{x} \,, \label{eq:Hwire}
\end{equation} 
where $\mu$ is the chemical potential, $\Delta$ the induced gap, and $q$ the Cooper pair momentum that models an externally applied supercurrent $I_w = e  v \hbar q$. 
We will consider here the zero temperature limit where many of the calculations can be done analytically. 
The Hamiltonian acts on the Nambu spinor  $\Psi(x)=(\psi_{\uparrow},\psi_{\downarrow},\psi_{\downarrow}^{\dagger},-\psi_{\uparrow}^{\dagger})^{T}$ with $ \tau_{x,y,z}  $ Pauli matrices acting on the  particle-hole space. 
 We note that the  spatial modulation of the pairing term gives rise to an effective Zeeman term $\tilde{V}_z = V_z + \frac{1}{2}v\hbar q $ by taking a unitary transformation into a moving frame $\tilde{H}(x,q)  = e^{-iqx\tau_z/2}H(x,q) e^{iqx\tau_z/2}$.
At the same time, the gap to quasiparticle excitations vanishes at a critical Zeeman energy $\tilde{V}_{z,c} = \pm \Delta$ which also gives the critical Cooper pair momentum $|q_c| = \Delta / (\frac{1}{2}\hbar v)$ at zero magnetic field~\cite{PhysRevB.107.224518}. 
At a non-zero magnetic field, the above expression predicts  critical Cooper pair momenta $q_c^{\pm} = (\pm \Delta - V_z)/(\frac{1}{2}\hbar v)$, with a gapped superconducting state in the interval $q_c^{-}(V_z) < q < q_c^{+}(V_z)$. Remarkably, the current created by $V_z$ opposite to $q$ enables a superconducting state even at large Zeeman field and applied currents as well as leads to a large diode effect on the edge, see Fig.~\ref{fig:Wire}c. 
The diode efficiency $\eta = ( |q_c^+| - |q_c^-|) / ( |q_c^+| + |q_c^-|)$ becomes 1 when $V_z = \pm \Delta$ and the critical current in one direction vanishes,  $q_c^{\pm} = 0$. 

The helicity of the helical edge can be reversed by setting $v \to -v$ in Eq.~(\ref{eq:Hwire}). 
In a system with two otherwise \textit{identical} topological insulator edges of opposite helicities, Fig.~\ref{fig:Wire}b, the diode effect vanishes since the combined system is inversion symmetric, Fig.~\ref{fig:Wire}c.  
By breaking the symmetry between the two edges one can recover the diode effect~\cite{PhysRevB.98.075430,PhysRevB.107.224518,PhysRevB.109.075412}, analogous to an asymmetric ring~\cite{burlakov2014superconducting}; In Fig.~\ref{fig:Wire}b we break the symmetry by assuming unequal induced gaps $\Delta_{1,2}$ for the two edges. 

The linearized description, Eq.~(\ref{eq:Hwire}), is valid near the Fermi level at energies much below the bulk band gap $E_g$ of the 2D topological insulator. 
One might therefore wonder if the large diode effect described above is an artifact of the linearized low-energy approximation. 
To answer this question, we simulate a 2D TI Hamiltonian in a ribbon geometry (Fig.~\ref{fig:Wire}b) by using a tight-binding model. 
Specifically, we discretize the BHZ model~\cite{doi:10.1126/science.1133734} using \texttt{Kwant}~\cite{Groth_2014} python package~\footnote{Our code to reproduce the figures is available at {https://purr.purdue.edu/publications/4493/1}.}.
While the original model~\cite{doi:10.1126/science.1133734} was for a HgTe quantum well, we use model parameters that make simulations computationally cheaper but still belong to the same topological phase. 
In our model, we consider two different chemical potentials $\mu=-0.149E_g$ and $-0.484E_g$ (measured from the Dirac point) corresponding to superconducting coherence lengths ($v_F$ denotes the Fermi velocity) $\hbar v_F / \Delta \approx 40a$ and $21a$, respectively, in units of the lattice constant $a$, and observe no qualitative difference in the results. In the remaining of this article, we show the results for the second case. 
The ribbon width $L_y=80a$  is chosen to be much larger than the edge penetration depth $\hbar v / E_g \approx 0.56a$, and so that the two edges do not couple to each other.

We find numerically the values of $q$ at which the gap to quasiparticle excitation closes. 
The resulting critical momenta $q_c^\pm$ vs $V_z$ are shown in Fig.~\ref{fig:Wire}c, where we consider three cases: (i) $\Delta_1=\Delta_2=\Delta$, (ii) $\Delta_1=\Delta$, $\Delta_2=5\Delta_1$, and (iii) $\Delta_1=\Delta/2$, $\Delta_2=5\Delta_1$, with $\Delta/E_g=0.0134$. In case (i), the system is inversion symmetric and shows no diode effect, while in case (ii) and (iii), we observe diode effects. 
To identify which edge becomes normal at $q=q_c$, we also 
investigate the probability densities of the zero-energy gapless quasiparticle state at the SC/N phase boundary. 
At boundaries with negative slopes in $V_z$-$q_c$ plane, the lower edge of the ribbon turns normal. The upper edge becomes normal at the boundaries with positive slopes in cases (i) and (iii). These observations are expected from the low-energy model, Eq.~(\ref{eq:Hwire}). 
However, in case (ii) which has a larger $\Delta$ than case (iii), we observe that the gapless quasiparticle states at the positive-slope boundary have a non-zero  probability density  in the bulk region of the ribbon, and as a consequence, the slope of the boundary is different from those due to edge states. 
This is not predicted by the lower-energy model. We note that for our choice of $\mu$, the valence band edge is only a distance $0.04 E_g \approx 3 \Delta$ from the Fermi level in this case.

\begin{figure*}[th]
    \centering
    \includegraphics{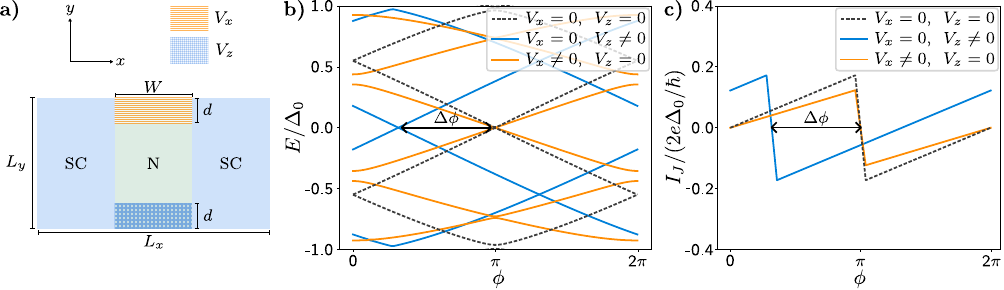}
    \caption{Josephson junction on a 2D TI. a) Geometry of the junction in our numerical model. Magnetic fields are applied to the edges of the normal region of the junction. b) Single-edge Andreev levels, obtained by setting $V_x=8\Delta_0,V_z=0.6\Delta_0$ in a). The dashed line shows the levels under zero field. A $V_x$ field on the edge flattens the levels (orange)  whereas a $V_z$ field shifts the levels (blue). c) Single-edge Josephson currents, which account for the total Josephson current calculated from b). A $V_x$ field suppresses the single-edge contribution whereas a $V_z$ field shifts the contribution in $\phi$.}  
    \label{fig:JJ Geometry and Effect of Bx Bz}
\end{figure*}

Next, we study non-reciprocity in a 2D TI Josephson junction, depicted in Fig.~\ref{fig:JJ Geometry and Effect of Bx Bz}a. 
We consider a junction of width $W$ where the two superconductors have a phase difference $\phi$. To break inversion and time-reversal symmetry we consider a magnetic field-induced  Zeeman energy  in the junction. 
The bottom half of the junction at $-\frac{W}{2} < x < \frac{W}{2}$ can be therefore described by the effective Hamiltonian~(\ref{eq:Hwire}) with the replacements 
 $V_z\sigma_z \to \Theta(\frac{W}{2} - |x|) V_i \sigma_i$ ($i=x,z$) and 
\begin{equation}\label{eq:H_SC,JJ}
   \Delta e^{iqx\tau_z}\tau_x \to  \text{Re}[\Delta(x)]\tau_x - \text{Im}[\Delta(x)]\tau_y,
\end{equation}
where $\Delta(x) = \Theta(|x| - \frac{W}{2}) \Delta_0 e^{i \phi(x)}$ and $\phi(x) = \Theta(x - \frac{W}{2}) \phi$.

For each value of the Josephson phase $\phi$, the system has a set of energy eigenvalues $E(\phi)$. The zero-temperature Josephson current of the junction is defined as~\cite{Tinkham2004} 
\begin{equation}\label{eq:I_J Definition}
    I_J(\phi) = \frac{2e}{\hbar}\frac{\text{d}E_\text{GS}(\phi)}{\text{d}\phi}.
\end{equation}
Here $E_\text{GS}$ is the ground state energy of the corresponding many-body system, which can be written as the sum of all negative energy eigenvalues $\{E^-_i(\phi)\}$ of our single-electron system:
\begin{eqnarray}\label{eq:EGS}
    E_\text{GS}(\phi) = \sum_{i=1}^\infty E_i^-(\phi).
\end{eqnarray}

We numerically study  $E_\text{GS}(\phi)$ and $I_J(\phi)$ of the junction depicted in Fig.~\ref{fig:JJ Geometry and Effect of Bx Bz}a. In our tight-binding simulation we model a 2D $L_x\times L_y$ system with  
$L_x=120a$, $L_y=80a$, $W=40a$, and $\xi = 21a$. 
The Zeeman fields in the junction are applied to rectangular regions of size $W\times d$ with $d = 10a$ chosen to be much larger than the edge penetration depth.

We calculated the Josephson current in various combinations of Zeeman potentials $V_z$ or $V_x$ applied to the two edges. 
Since $L_y$ is much larger than the edge penetration depth, the current can be understood as the sum of two contributions, from the upper and lower edge:
\begin{equation}\label{eq:I_J as a sum of upper and lower}
    I_J(\phi) = I_{J,\text{upper}}(\phi) + I_{J,\text{lower}}(\phi),
\end{equation}
and the effect of applying magnetic fields to the edges on $I_J(\phi)$ can be understood in terms of these single-edge contributions.

We consider  the limit of a long junction $W \gtrsim \xi$, where the Josephson current  $I_J(\phi)$ is a piecewise linear function~\cite{PhysRevLett.110.017003} as are its single-edge contributions in Eq.~(\ref{eq:I_J as a sum of upper and lower}). We can model each term with a $2\pi$-periodic piece-wise linear function~\cite{PhysRevB.92.035428}:
\begin{equation}\label{eq:f(phi)}
    I_{J,\text{single edge}}(\phi) \approx
    \begin{cases}
        \frac{I_0}{\pi - t}\phi & \text{if } \phi \bmod 2\pi \in[0, \pi - t)\\
        -\frac{I_0}{t}(\phi - \pi) & \text{if } \phi\bmod 2\pi \in [\pi - t, \pi + t)\\
        \frac{I_0}{\pi - t}(\phi - 2\pi) & \text{if } \phi\bmod 2\pi \in [\pi + t, 2\pi),
    \end{cases}
\end{equation}
where $I_0$ is the maximum single edge Josephson current and $\pi \pm t$ are the $\phi$ values at which $I_{J,\text{single edge}}(\phi)$ reaches minimum and maximum, respectively. 
Matching Eq.~(\ref{eq:f(phi)}) with our numerically calculated Josephson currents, we find $t \approx 0.0659$.

Next, we describe the effects of $V_x$ and $V_z$ on a single edge Josephson current. 
Applying a magnetic field $V_x$ perpendicular to the spin quantization axis on an edge gaps the Dirac point (Fig.~\ref{fig:Wire}a) and lowers the transmission of the edge state~\cite{PhysRevB.79.161408}. 
Thus, $V_x$ will have the effect of flattening the phase dispersion of Andreev states and thus suppressing the Josephson current, see Fig.~\ref{fig:JJ Geometry and Effect of Bx Bz}b-c. 
 
On the other hand, applying a $V_z$ field  on an edge only shifts the Andreev state energies in $\phi$ and consequently the Josephson current is shifted in $\phi$ while its maximum value does not change (Fig.~\ref{fig:JJ Geometry and Effect of Bx Bz}b-c). 
Thus, with $V_z$ one can realize a $\phi_0$ junction~\cite{PhysRevB.92.035428}. 
In particular, we find that for  values of $V_z$ where linearized model is valid ($V_z<|E_g - \mu|$), 
the shift $\Delta\phi$ of these curves in $\phi$ is linear, 
\begin{equation}\label{eq:Delta phi linear in b_z}
    \Delta\phi = c\frac{V_z}{\Delta_0}\text{ for }V_z< |E_g - \mu|, 
\end{equation}
where $c$ is a unit-free constant, estimated below, see Eq.~(\ref{eq:c estimate}). 
Since the Josephson current of a long junction is (piecewise) linear in $\phi$, the result Eq.~(\ref{eq:Delta phi linear in b_z}) is analogous to the Zeeman induced current on the helical edge, discussed below Eq.~(\ref{eq:HwireNormal}), 
with the exception that the Josephson current is a $2\pi$-periodic function of $\phi$.

The prefactor $c$ in Eq.~(\ref{eq:Delta phi linear in b_z}) can be estimated based on the low-energy model~(\ref{eq:Hwire}). The energy of the lowest Andreev bound state on the lower edge under zero field is~\cite{PhysRevB.92.134508}
\begin{equation}\label{eq:E(phi)}
    E_1(\phi) = \frac{1}{2} E_T \frac{W}{W + \xi}|\phi - \pi|,
\end{equation}
which, upon taking  the Thouless energy $E_\text{T}=\hbar v/W \approx 0.525 \Delta_0$ and $W = 1.904\xi$ (so that $E_1(\phi = 0) = 0.541\Delta_0$) matches well with the numerically calculated result, see Fig.~\ref{fig:JJ Geometry and Effect of Bx Bz}b. 
Higher-energy Andreev states have similar expressions~\cite{PhysRevB.92.134508}. 
The Andreev state is a $\sigma_z = \sigma  = \pm 1$ eigenstate whose wave function  extends a distance $\xi$ into the superconducting banks.  
Therefore, a Zeeman field $V_z$ in the junction shifts its energy by approximately $V_z \times W/(W+\xi) $ which in Eq.~(\ref{eq:E(phi)})  can be compensated by shifting $\phi$ by the amount given in Eq.~(\ref{eq:Delta phi linear in b_z}) with a proportionality constant
\begin{equation}
    c \approx  \sigma \frac{2W}{\xi} \approx 3.81 \sigma,  
    \label{eq:c estimate}
\end{equation}
which is on the same order of magnitude as a numerically obtained value of $3.57 \pm 0.03$. 
Here $\sigma = \pm 1$ depending on the helicity of the edge. Thus the shift $\Delta \phi$, Eq.~(\ref{eq:Delta phi linear in b_z}), is opposite on two edges of a junction with same $V_z$.

Next, we will investigate the effect of combined $V_x, V_z$ terms on the two edges of the junction, focusing on the Josephson diode effect.  
We define Josephson  diode efficiency 
\begin{equation} \label{eq:etaJ}
\eta_J = |I_{J,c}^{+}+I_{J,c}^{-}|/|I_{J,c}^{+}-I_{J,c}^{-}|  ,
\end{equation}
via the maximum and minimum Josephson currents (critical currents),
respectively $I_{J,c}^{+} = \max_{\phi}{I_J}$ and $I_{J,c}^{-} = \min_{\phi}{I_J}$, 
obtained from Eq.~(\ref{eq:I_J as a sum of upper and lower}).

\begin{figure*}[t]
    \centering
    \includegraphics{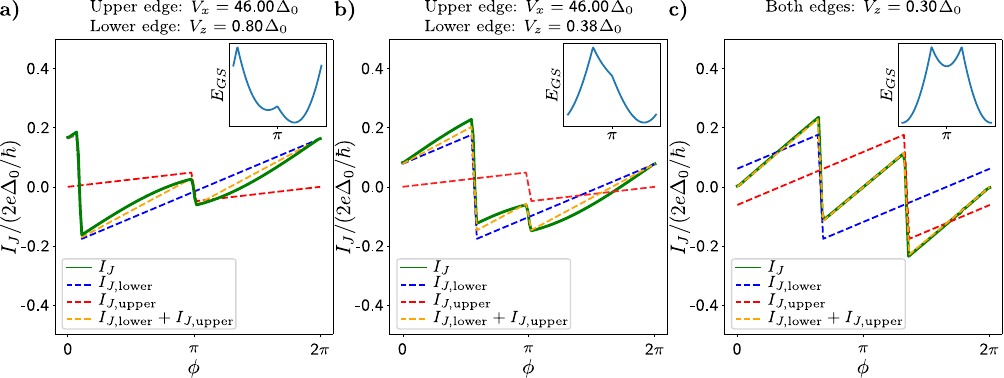}
    \caption{Josephson current and ground state energy (Josephson potential) of a 2D TI junction with Zeeman fields on the edges. The blue and red lines show the lower and upper edge current modeled by Eq.~(\ref{eq:f(phi)}), with an appropriate amount of $\phi$-shift and suppression so that their sum (yellow lines) matches the result calculated from Eqs.~(\ref{eq:I_J Definition}), (\ref{eq:EGS}) (green line). The ground state energies calculated from Eq.~(\ref{eq:EGS}) are shown as insets.
    a) $V_x$ on the upper edge and $V_z$ on the lower edge, with two stable minima in the ground state energy. If one is able to initialize the phase of the junction at either stable minima, the system will have two critical currents with a large difference. b) $V_x$ on the upper edge and $V_z$ on the lower edge, with one stable minimum in the ground state energy. The system exhibits diode effect. The discrepancy between the linear approximation and the tight-binding result is due to the curvature of the exact $I_{J,\text{single, edge}}(\phi)$ curve, which is eliminated as we move to a longer junction. c) Symmetric $V_z$ on both edges, where the system exhibits no diode effect. The ground state energy has two inequivalent but symmetric minima. }
    \label{fig:JJ Examples}
\end{figure*}

When the junction is inversion symmetric, such as in the case of 
applying a symmetric $V_z$ field on both edges, we have $\eta_J = 0$. By applying a symmetric $V_z$ on both edges, the $E_i(\phi)$ curves of the upper and lower edge states and their contributions to the Josephson current are shifted by the same amount in $\phi$, but in opposite directions, see Fig.~\ref{fig:JJ Geometry and Effect of Bx Bz}b-c and Eq.~(\ref{eq:c estimate}). 
The ground state energy (Josephson potential) has two inequivalent minima, but each minimum is symmetric and therefore the critical current shows no diode effect, see Fig.~\ref{fig:JJ Examples}c~\footnote{The critical current can however~\cite{PhysRevB.76.224523} differ significantly, depending on which minimum the system is initialized in. }. 
Asymmetrically applied $V_z$ can be symmetrized by an appropriate shift of $\phi$, without changing the Josephson critical current. 
Thus, there is no diode effect in a long junction with only $V_z$, even if applied  asymmetrically. 
This agrees with small magnetization limit of Ref.~\onlinecite{PhysRevB.109.075412}. 
However, they found a non-zero diode effect in a short junction with large magnetization where Eq.~(\ref{eq:Delta phi linear in b_z}) might not hold.

By applying $V_x$ on one edge, we can break inversion symmetry. However, without $V_z$, the Josephson current~(\ref{eq:I_J as a sum of upper and lower}) can be again symmetrized and will have no diode effect. 
Diode effect requires both $V_x$ and $V_z$ in an inversion-symmetry breaking way. 
Such combination leads to an asymmetric Josephson potential and non-reciprocal critical currents, see  Fig.~\ref{fig:JJ Examples}a-b. 
With certain combinations of $V_x$ and $V_z$, the Josephson potential will have two stable minima, see Fig.~\ref{fig:JJ Examples}a. If one can initialize the phase of the junction into one of the stable minima $\phi_*$, the effective critical currents of the system will be the two  extrema of the Josephson current nearest to $\phi_*$~\cite{PhysRevB.76.224523,2023Sci...382.1422Z}. The difference between the two critical currents in such cases can be large, as in Fig.~\ref{fig:JJ Examples}a, even if the ``global'' diode efficiency~(\ref{eq:etaJ}) is small. 

\begin{figure}
    \centering
    \includegraphics[width=\columnwidth]{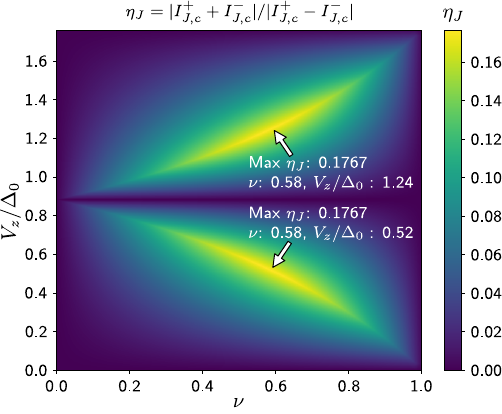}
    \caption{The Josephson diode efficiency calculated from Eqs.~(\ref{eq:I_J as a sum of upper and lower}), (\ref{eq:f(phi)}), (\ref{eq:etaJ}), where we take $t=0.0659$. 
    We assume that Zeeman fields $V_x,V_z$ are applied to top and bottom edges, respectively (see Fig.~\ref{fig:JJ Geometry and Effect of Bx Bz}a). 
    The top edge critical current $I_0(V_x)$ is controlled by $V_x$, parametrized on  the horizontal axis by  
    $\nu=1-I_0(V_x)/I_0(0)$. 
    The vertical axis shows $V_z$ applied to the  bottom edge, leading to a shift~(\ref{eq:Delta phi linear in b_z}) to its Josephson current; we take  $c=3.57$. 
    The figure is $2\pi$-periodic in $c V_z / \Delta_0$. 
    In the limit $t\ll 1$, the analytically calculated maximum diode efficiency is $\max{\eta_J}=(\sqrt{2}-1)^2\approx 0.17$, with $\nu=2-\sqrt{2}\approx 0.59$ and 
    $|cV_z/\Delta_0|=(2-\sqrt{2})\pi,\sqrt{2}\pi$, respectively.}
    \label{fig:JJ efficiency}
\end{figure}

For a zero-temperature long Josephson junction, where Eqs.~(\ref{eq:I_J as a sum of upper and lower}) and (\ref{eq:f(phi)}) hold, the diode efficiency~(\ref{eq:etaJ}) can be calculated analytically. 
In Fig.~\ref{fig:JJ efficiency} we show the effect of $V_x$ on one edge and $V_z$ on the other on the diode efficiency $\eta_J$.

In summary, we have shown that by applying Zeeman fields to the edges of a 2D TI, one can achieve a large non-reciprocal critical current. Specifically, for a uniform 2D TI edge, the maximum diode efficiency is 1. For a zero-temperature long Josephson junction, the maximum diode efficiency is $(\sqrt{2}-1)^2 \approx 0.17$, see Fig.~\ref{fig:JJ efficiency}. 
An open question is how one can obtain the maximum diode efficiency for a finite-temperature junction of a general length with Zeeman fields on the edges, and the corresponding optimal temperature, junction length, and Zeeman fields; Our preliminary results show that the diode efficiency is generally increased at non-zero temperature. 
A second open question is whether one can utilize the asymmetric Josephson potential with two minima to obtain highly non-reciprocal critical currents not solely characterized by $\eta_J$. 

Upon finalizing this article, we learned of Ref.~\onlinecite{2024arXiv240317894F} where a short 2D TI Josephson junction is studied. 
Other than differences between short and long (our case) junctions, the results seem to agree where there is overlap.

\begin{acknowledgments}
JIV thanks Michael Fuhrer for encouragement and  Yuli Lyanda-Geller and Leonid Rokhinson  for helpful discussions. 
This material is based upon work supported by the U.S. Department of Energy, Office of Science, National Quantum Information Science Research Centers, Quantum Science Center. 
\end{acknowledgments}


\providecommand{\noopsort}[1]{}\providecommand{\singleletter}[1]{#1}

\end{document}